\documentclass[11pt,
showpacs,
preprintnumbers,amsmath,amssymb,
aps,prd,nofootinbib,eqsecnum,a4paper]{revtex4}
\usepackage{epsfig}
\usepackage{graphicx,epsf}
\usepackage{color}
\usepackage{bm}
\usepackage{psfrag}
\usepackage[symbol]{footmisc}
\usepackage{tensor}
\usepackage{hyperref}
\def\beq{\begin{equation}}
\def\eeq{\end{equation}}
\def\bea{\begin{eqnarray}}
\def\eea{\end{eqnarray}}
\def\bi{\begin{itemize}}
\def\ei{\end{itemize}}
\def\cs2{c_{\rm{s}}^2}
\def \beg {\begin{enumerate}}
\def \en {\end{enumerate}}

\def\M0{{\cal M}_0}

\DeclareMathOperator{\sech}{sech}
\begin{document}

\title{Constraints for the running index independent of the parameters of the model}

\author{Gabriel Germ\'an\footnote[1]{\href{mailto:gabriel@icf.unam.mx}{gabriel@icf.unam.mx}}}
\affiliation{
$^2$Instituto de Ciencias F\'{i}sicas, Universidad Nacional Aut\'{o}noma de M\'{e}xico,\\Av. Universidad S/N. Cuernavaca, Morelos, 62251, M\'{e}xico}
%

\begin{abstract}
  By writing the running of the scalar spectral index completely in terms of the scalar index $n_s$ and the tensor-to-scalar ratio $r$ we are able to impose constraints to models of inflation which are independent of the parameters of the model in question. We write analytical expressions for the running index of Natural Inflation, two models of the type Mutated Hilltop Inflation and the Starobinsky model. The resulting  formulae for the running  depend exclusively on $n_s$ and/or $r$ and will keep tightening the running index further as additional conditions and observations constrain the scalar and the tensor-to-scalar indices.
  
\end{abstract}


\maketitle

\section {\bf Introduction}\label{Intro}
The inflationary paradigm \cite{Starobinsky:1980te}, \cite{Guth:1980zm}, has been introduced some forty years ago in order to solve some important problems of the old Big-Bang cosmology. While such a solution is compelling and attractive it does not seem to require a specific model of inflation with very particular characteristics, for this reason even now we do not yet have a definitive model (for reviews see eg, \cite{Linde:1984ir}-\cite{Martin:2018ycu}). Various models are able to satisfy the available data and distinguish themselves from others by their construction and physical motivation, for this reason it is important to establish model-independent results which can help to discriminate among the plethora of existing viable models \cite{Martin:2013tda}.  At least, to establish general results which are independent of the particular characteristics of each model.

The purpose of this work is to obtain bounds for the running of the scalar spectral index for several models of interest, but with the bounds nevertheless independent of the parameters of the model in question. For it we express the running of the tensor and the scalar spectral indices in terms of the scalar index $n_s$ and/or the tensor-to-scalar ratio $r$. While the resulting expression is clearly particular to the model under consideration it does not involve any of the parameters of such model and  the phenomenological values of the observables $n_s$ and $r$ are directly used to constrain the running. This is done for four specific models, once the analytical formula for the running is obtained it is easy to get the bounds as dictated by the range of values for $n_s$ and $r$ provided by the latest results from the Planck Collaboration. We also discuss the possibility of breaking degeneracies amongst the models by using the running index. 

The outline of the paper is as follows: in section \ref{THE} we give general results which will be used in the subsequent sections. We also establish a formula for the running of the tensor spectral index which is model independent and should be satisfied by any single field model of inflation. A simple but general formula for the slow-roll parameter $\eta$ implying a downward concave potential is given. In sections~\ref{NI} to \ref{STA} we obtain the running for several models, find their respective bounds and discuss some important features for each model under study. Finally, Section~\ref{CON} contains the main conclusions of the paper.
\section {\bf The general approach}\label{THE} 

The connection with inflation-based models is made initially through the primordial power spectra {$\mathcal{P}_i$} parameterized by a power law of scalar and tensorial perturbations. These are generally given in terms of the spectral amplitude $ A_i $ together with the spectral indices $ n_i $, where the subscript $i$ refers to scalar $(s)$ or tensor $(t)$ components (see e.g., \cite{Ade:2015lrj})
\begin{eqnarray}
\mathcal{P}_s(k)&=&A_s \left( \frac{k}{k_{k_p}}\right)^{(n_s-1) },\label{ps1}\\
\mathcal{P}_t(k)&=&A_t \left( \frac{k}{k_{k_p}}\right)^{n_t}=rA_s\left( \frac{k}{k_{k_p}}\right)^{n_t}.\label{pt1}
\end{eqnarray}
Here $k$ is the wave number mode and $r\equiv \mathcal{P}_t(k)/\mathcal{P}_s(k)$ the ratio of tensor-to-scalar perturbations at
the pivot scale $k=k_p$\footnote{The subindex $k$ or $k_p$ above denotes the value of the inflaton when scales the size of the pivot scale leave the horizon.} .
Slow-roll (SR) inflation predicts the spectrum of curvature perturbations to be close to scale-invariant. This allows a simpler parametrization of the spectra in terms of quantities evaluated at $k_p$  such as the spectral indices and the running of scalar and tensor perturbations (see e.g., \cite{Ade:2015lrj})
\begin{eqnarray}
\mathcal{P}_s(k)&=&A_s \left( \frac{k}{k_{k_p}}\right)^{(n_s-1) + \frac{1}{2}n_{sk}  \ln\left(\frac{k}{k_{k_p}}\right)} ,\label{ps2}\\
\mathcal{P}_t(k)&=&A_t \left( \frac{k}{k_{k_p}}\right)^{n_t+ \frac{1}{2} n_{tk} \ln\left(\frac{k}{k_{k_p}}\right) }\label{pt2},
\end{eqnarray}
where $n_{sk} \equiv \frac{d n_{s}}{d \ln k}$ is the running of the scalar index $n_{s}$ and $n_{tk}\equiv \frac{d n_{t}}{d \ln k}$ the running of the tensor spectral index $n_{t}$,  in a self-explanatory notation. In the literature $n_{sk}$ is usually denoted by $\alpha$ but here we prefer to use this more symmetrical notation between scalar and tensorial quantities.
Contact with models of inflation is achieved precisely through these indices (also called observables) which in the SR approximation (first introduced in the context of a bouncing cosmology with two quasi-de Sitter stages \cite{Starobinsky:19780}) are given by (see e.g.,  \cite{Lyth:1998xn}, \cite{Liddle:1994dx})
\begin{eqnarray}
n_{t} &=&-2\epsilon = -\frac{r}{8} , \label{Int} \\
n_{s} &=&1+2\eta -6\epsilon ,  \label{Ins} \\
n_{tk} &=&4\epsilon\left( \eta -2\epsilon\right), \label{Intk} \\
n_{sk} &=&16\epsilon \eta -24\epsilon ^{2}-2\xi_2, \label{Insk} \\
A_s(k) &=&\frac{1}{24\pi ^{2}} \frac{\Lambda^4}{%
\epsilon}, \label{IA} 
\end{eqnarray}
where $A_s(k)$ is the amplitude of density perturbations at wave number $k$ and $\Lambda$ is the scale of 
inflation, with $\Lambda \equiv V_{k}^{1/4}$. 
The slow-roll parameters appearing above are 
\begin{equation}
\epsilon \equiv \frac{M^{2}}{2}\left( \frac{V^{\prime }}{V }\right) ^{2},\quad\quad
\eta \equiv M^{2}\frac{V^{\prime \prime }}{V}, \quad\quad
\xi_2 \equiv M^{4}\frac{V^{\prime }V^{\prime \prime \prime }}{V^{2}},
\label{Spa}
\end{equation}
and should be evaluated at $k$. Also,  $M$ is the reduced Planck mass $M=2.44\times 10^{18} \,\mathrm{GeV}$ which we set equal 
to one in what follows, primes on $V$ denote derivatives with respect to the inflaton $\phi$.

In general, a model independent constrain among observables results from Eqs.~\eqref{Int} to \eqref{Intk}, \cite{Carrillo-Gonzalez:2014tia}
\begin{equation}
n_{tk} =\frac{1}{64}r\left( r-8 \delta_{n_s}\right) \;,
\label{ntkcons}
\end{equation}
where $\delta_{n_s}$ is defined as $\delta_{n_s}\equiv 1-n_s$. From the range for the spectral index $n_s=0.9649\pm 0.0042$ and tensor-to-scalar ratio $r < 0.063$ reported by the Planck collaboration \cite{Aghanim:2018eyx}, \cite{Akrami:2018odb}, $n_{tk}$ is bounded as follows 
\begin{equation}
 -2.45\times 10^{-4} < n_{tk} < 0\;.
\label{Intkbounds}
\end{equation}
\begin{figure}[tb]
\includegraphics[width=10cm]{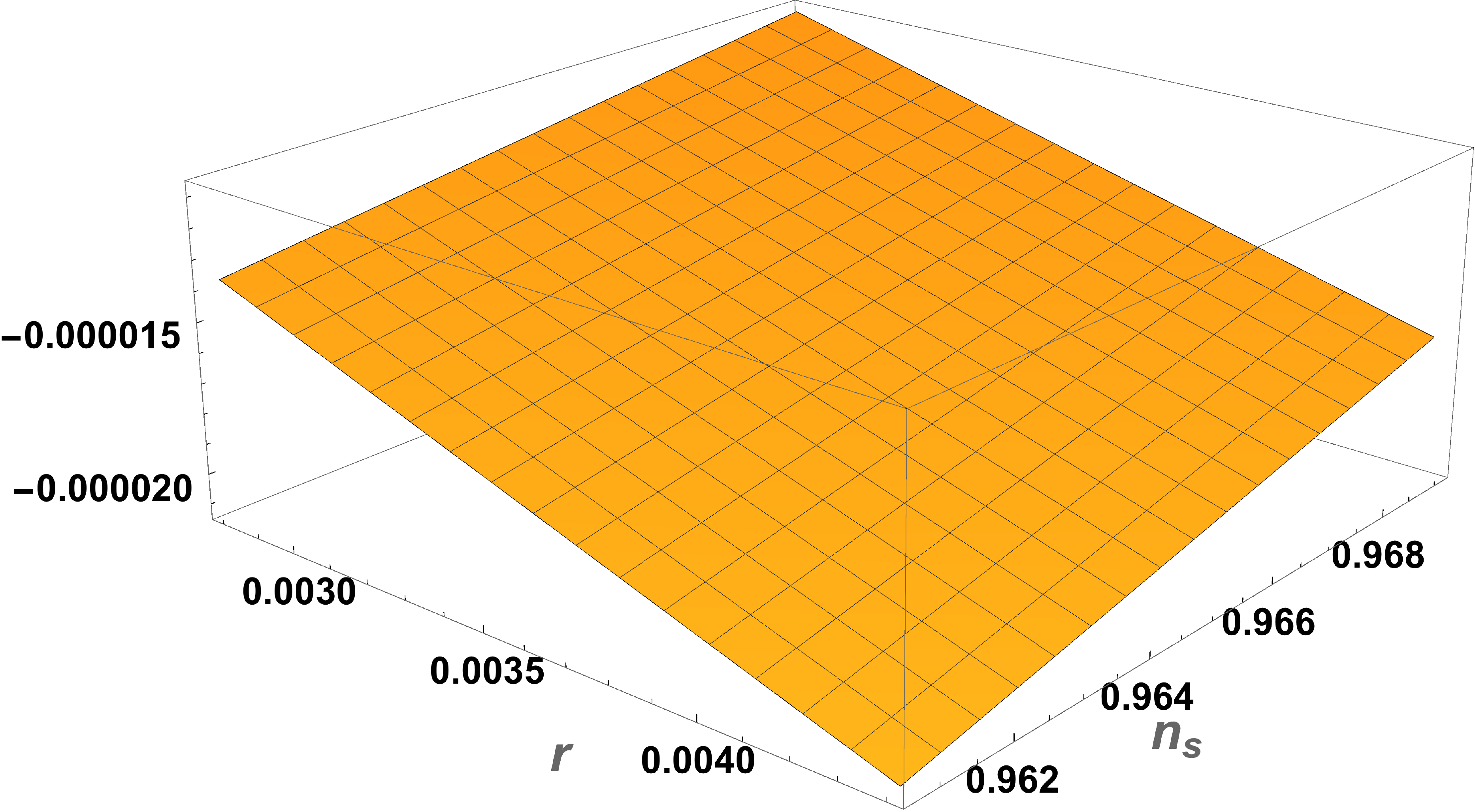}
\caption{\small Plot of the running of the tensor spectral index $n_{tk}$ given by Eq.~\eqref{ntkcons} as a function of the tensor-to-scalar ratio $r$ and the scalar spectral index $n_s$. From the range for $n_s=0.9649\pm 0.0042$ and  $r < 0.063$ reported by the Planck collaboration \cite{Aghanim:2018eyx}, \cite{Akrami:2018odb}, Eq.~\eqref{ntkcons} is is bounded as  follows $-2.45\times 10^{-4} < n_{tk} < 0.$ Both Eq.~\eqref{ntkcons}  and its bounds are model independent and any single field inflationary model should satisfy them.
}
\label{ntk}
\end{figure}

Also, from Eqs.~\eqref{Int} and \eqref{Ins} we find that
\begin{equation}
\eta =\frac{1}{16}\left( 3 r-8 \delta_{n_s}\right) \;.
\label{Ieta}
\end{equation}
 From the bounds for $n_s$ and $r$ given above, $\eta$ is bounded as $-0.01965 < \eta < -0.00364$ thus, a downward concave potential is preferred. Note that Eqs.~\eqref{ntkcons} and \eqref{Ieta} and their corresponding bounds are model independent and should be satisfied by any single field model of inflation.

Using Eqs.~\eqref{Ieta} and \eqref{Int}, the expression for the running of the scalar index given by Eq.~\eqref{Insk} can be written as
\begin{equation}
n_{sk} =\frac{3}{32}r^2 - \frac{1}{2} \delta_{n_s} r -  \frac{1}{4} r \frac{V^{\prime \prime \prime }}{V^{\prime}} \;,
\label{nsk}
\end{equation}
this is as far as we can get writing $n_{sk}$ in terms of $n_s$ and $r$ in a model independent way. The exercise which follows consists in finding $V^{\prime \prime \prime }/V^{\prime}$ in terms of $n_s$ and $r$. In this way we will have an expression for $n_{sk}$, specific for the model in question, but independent of the parameters of such model. Thus, the bounds on $n_{sk}$ will be  obtained directly from the bounds for the observables $n_s$ and $r$ without specifying any particular value for the parameters of the model in question. In what follows we find $n_{sk}=n_{sk}(n_s,r)$ and its corresponding bounds for four models: Natural Inflation, two models of the type Mutated Hilltop Inflation and the Starobinsky model.
\section {\bf Natural inflation}\label{NI}
\begin{figure}[tb]
\includegraphics[width=10cm]{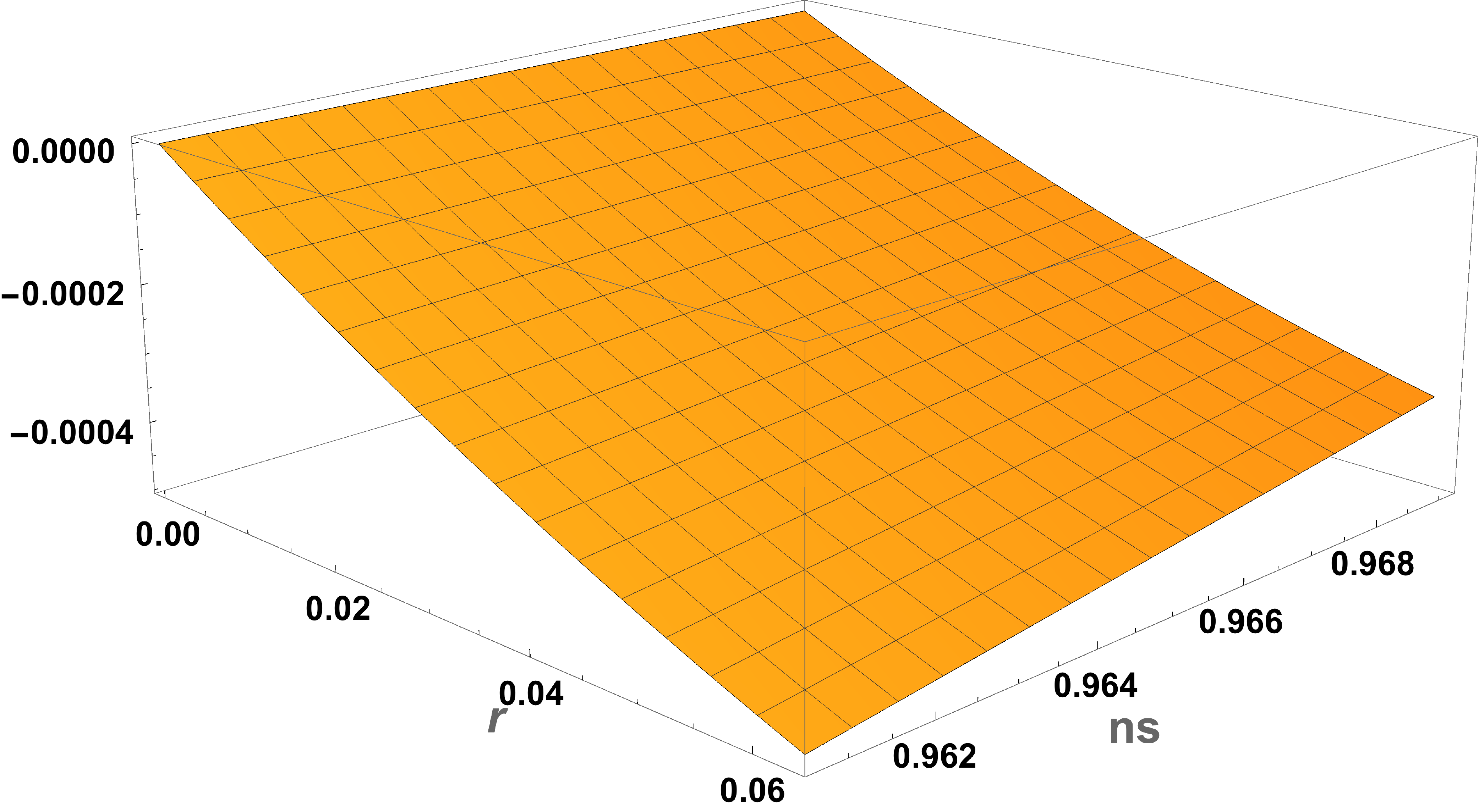}
\caption{\small Plot of the running of the scalar index given by Eq.~\eqref{nsksol} for the NI model defined by the potential Eq.~(\ref{nipot}). The Plotted running is certainly model dependent but does not depends on the parameters of the model itself. The bounds on $n_{sk}$, $-4.9\times 10^{-4} < n_{sk} < 0$ depend exclusively on the phenomenological bounds for the spectral index $n_s$ and tensor-to-scalar ratio $r$ as reported by the Planck collaboration \cite{Aghanim:2018eyx}, \cite{Akrami:2018odb}.
}
\label{nsknifig}
\end{figure}

\begin{figure}[tb]
\includegraphics[width=10cm]{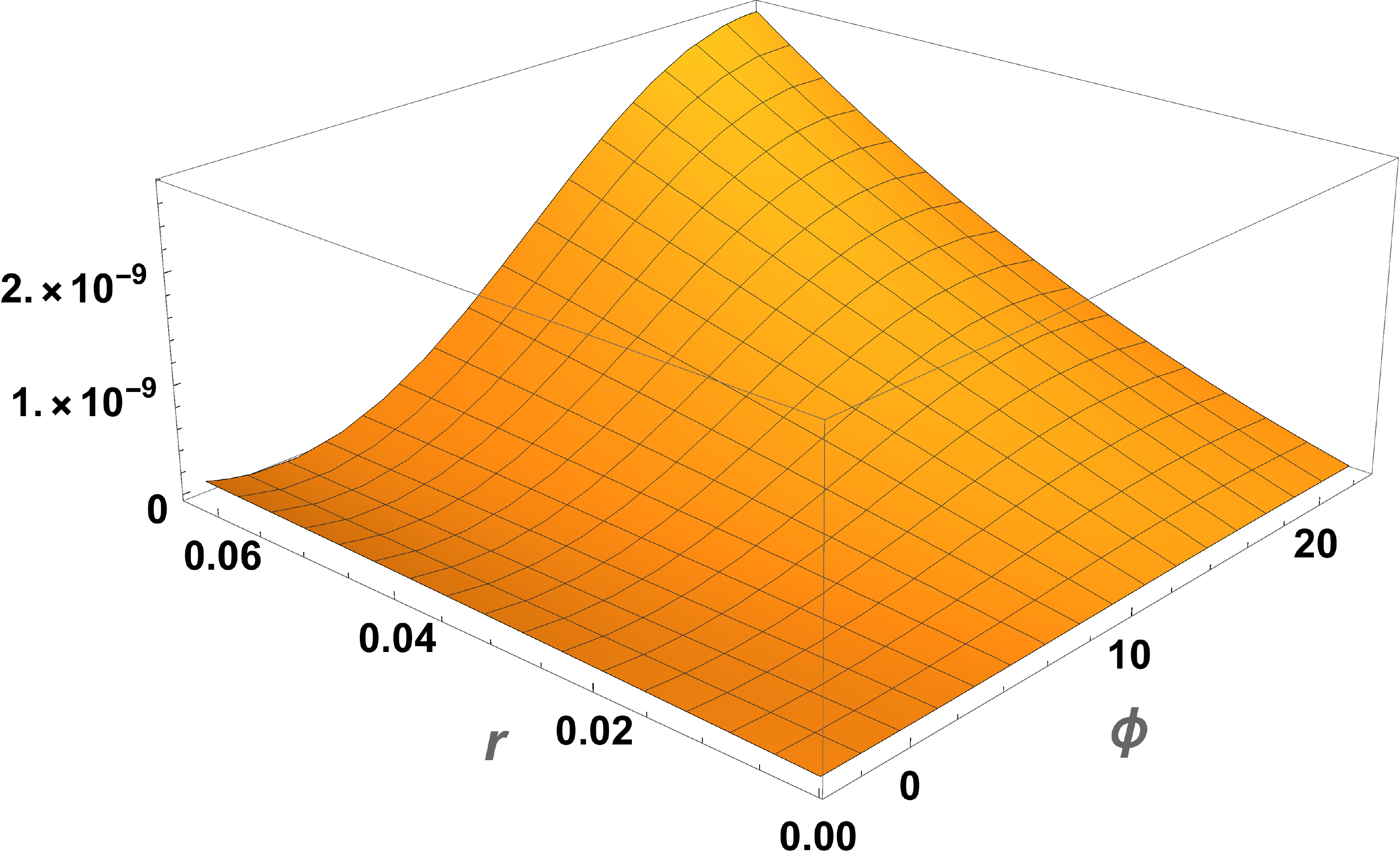}
\caption{\small The Natural Inflation potential given by Eq.~(\ref{potnidos})  as a function of $\phi$ and of the tensor-to-scalar ratio $r$ for the central value $n_s=0.9649$. As $r$ goes to cero so does the scale of inflation $\Lambda\equiv V_k^{1/4}$, being proportional to $r^{1/4}$.
}
\label{potni2}
\end{figure}

The potential for Natural inflation (NI) is  \cite{Freese:1990rb}, \cite{Adams:1992bn}
\begin{equation}
V =V_0\left(1-\cos(\frac{\phi}{f})\right)\;,
\label{nipot}
\end{equation}
this is a two-parameter model however, by working with Eqs.~\eqref{Int} and \eqref{Ins} we only have to deal with $f$. 
\begin{figure}[tb]
\includegraphics[width=8cm]{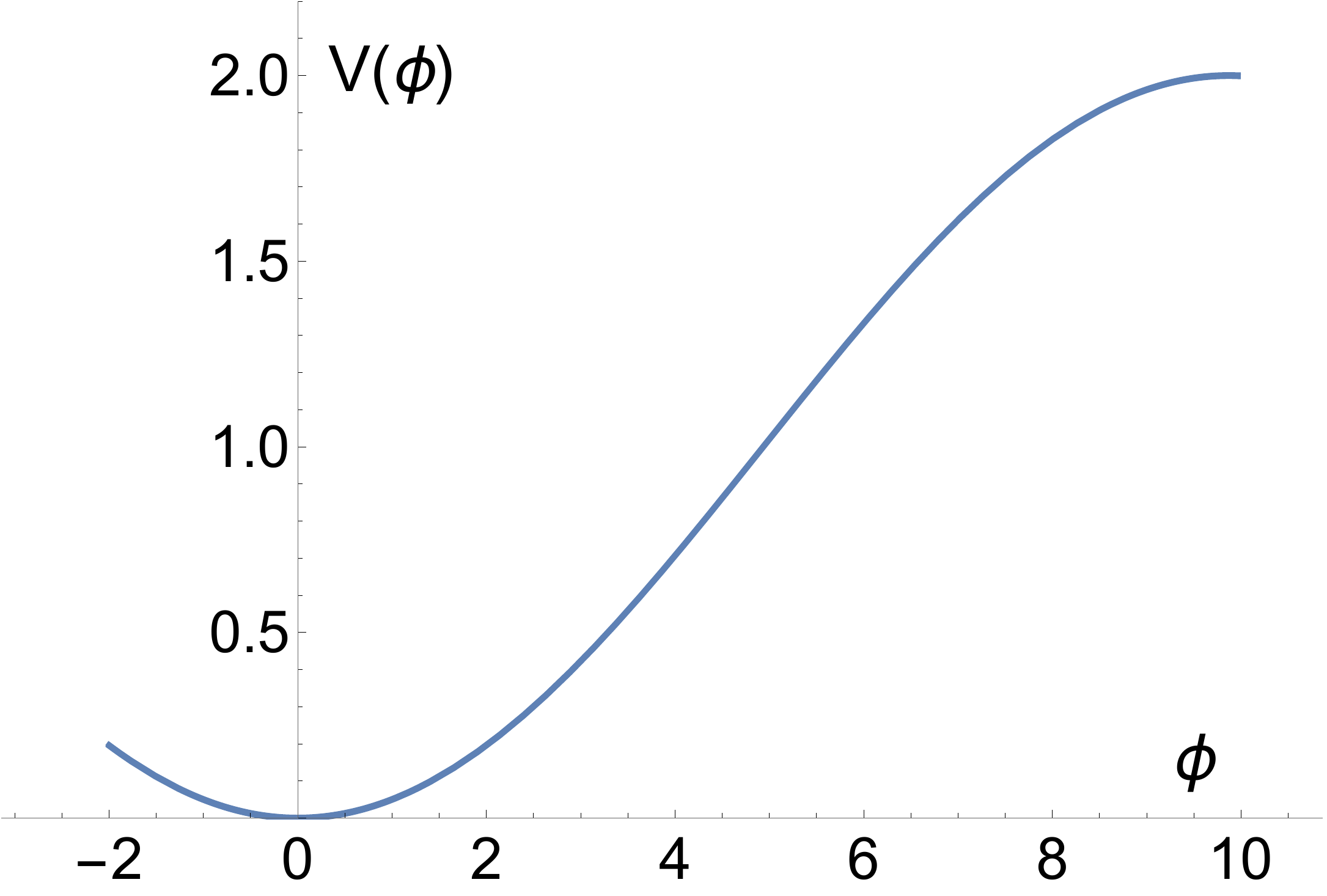}
\caption{\small Schematic plot of the Natural Inflation (NI) potential given by Eq.~(\ref{nipot})  as a function of $\phi$ for an inflaton field rolling from the right. }
\label{potni}
\end{figure}
From the expression $r=16\epsilon$, which for NI can be written as
\begin{equation}
r - \frac{8\sin^2(\frac{\phi_k}{f})}{f^2\left(1 - \cos(\frac{\phi_k}{f})\right)^2} = 0\;,
\label{req}
\end{equation}
we get
\begin{equation}
\cos(\frac{\phi_k}{f}) = 1-\frac{16}{8+f^2\, r}\;.
\label{reqsol}
\end{equation}
Evaluating  $\eta$ with this solution we find that Eq.~\eqref{Ins} 
\begin{equation}
\delta_{n_s} + \frac{1}{8}\left(r-\frac{8}{f^2}\right) - \frac{3}{8}r= 0\;,
\label{nseq}
\end{equation}
($\delta_{n_s}\equiv 1-n_s$) becomes an equation for $f=f(n_s,r)$  with the solution 
\begin{equation}
f =  \frac{2}{\sqrt{4 \delta_{n_s}-r}}\;.
\label{fsol}
\end{equation}
Thus,
\begin{equation}
\frac{V^{\prime \prime \prime }}{V^{\prime}}=  -\frac{1}{f^2} = - \delta_{n_s}  + \frac{1}{4}r\;,
\label{v3vpni}
\end{equation}
this last result together with Eq.~\eqref{nsk}  implies
\begin{equation}
n_{sk} = \frac{1}{32}r\left(r-8 \delta_{n_s}\right) \;,
\label{nsksol}
\end{equation}
(see Eqs.~(11) and (13) in Ref.~\cite{Carrillo-Gonzalez:2014tia}). Comparing Eq.~\eqref{nsksol} with \eqref{ntkcons} we see that, for this model, $n_{sk} = 2\, n_{tk}$.
From the bounds for $n_s$ and $r$, $n_{sk}$ is bounded as follows (see Fig.~(\ref{nsknifig}))
\begin{equation}
0.9607<n_s<0.9691, \quad  0.063 > r > 0, \quad \Rightarrow  \quad -4.9\times 10^{-4}<n_{sk}<0\;.
\label{nibounds}
\end{equation}
Finally, from Eq.~\eqref{IA} we can find $V_0$ in terms on $n_s$ and $r$ with the result $V_0=\frac{3A_s\pi^2r(8\delta_{n_s}-r)}{8(4\delta_{n_s}-r)}$ which together with Eq.~\eqref{fsol} allows to rewrite the potential as
\begin{equation}
V(\phi)=\frac{3A_s\pi^2r(8\delta_{n_s}-r)}{4(4\delta_{n_s}-r)}\sin^2\left(\frac{\sqrt{4\delta_{n_s}-r}}{4}\phi\right)\;,
\label{potnidos}
\end{equation}
of course $\phi$ is not an observable neither is the potential. In Fig.~(\ref{potni2})) we show the potential as a function of $\phi$ and the tensor-to-scalar ratio $r$ for $n_s$ fixed to the central value of the Planck range $n_s=0.9649$. As $r$ takes smaller values so does $V$ (an the inflationary energy scale) as expected from Eq.~\eqref{IA} written in the form $V=\frac{3\pi^2 A_s}{2}r$. The potential keeps its shape for any slice in the $V$ vs $\phi$ plane but its height is lowered by $r$.
\section {\bf Mutated Hilltop Inflation}\label{MUT}
\begin{figure}[tb]
\includegraphics[width=8cm]{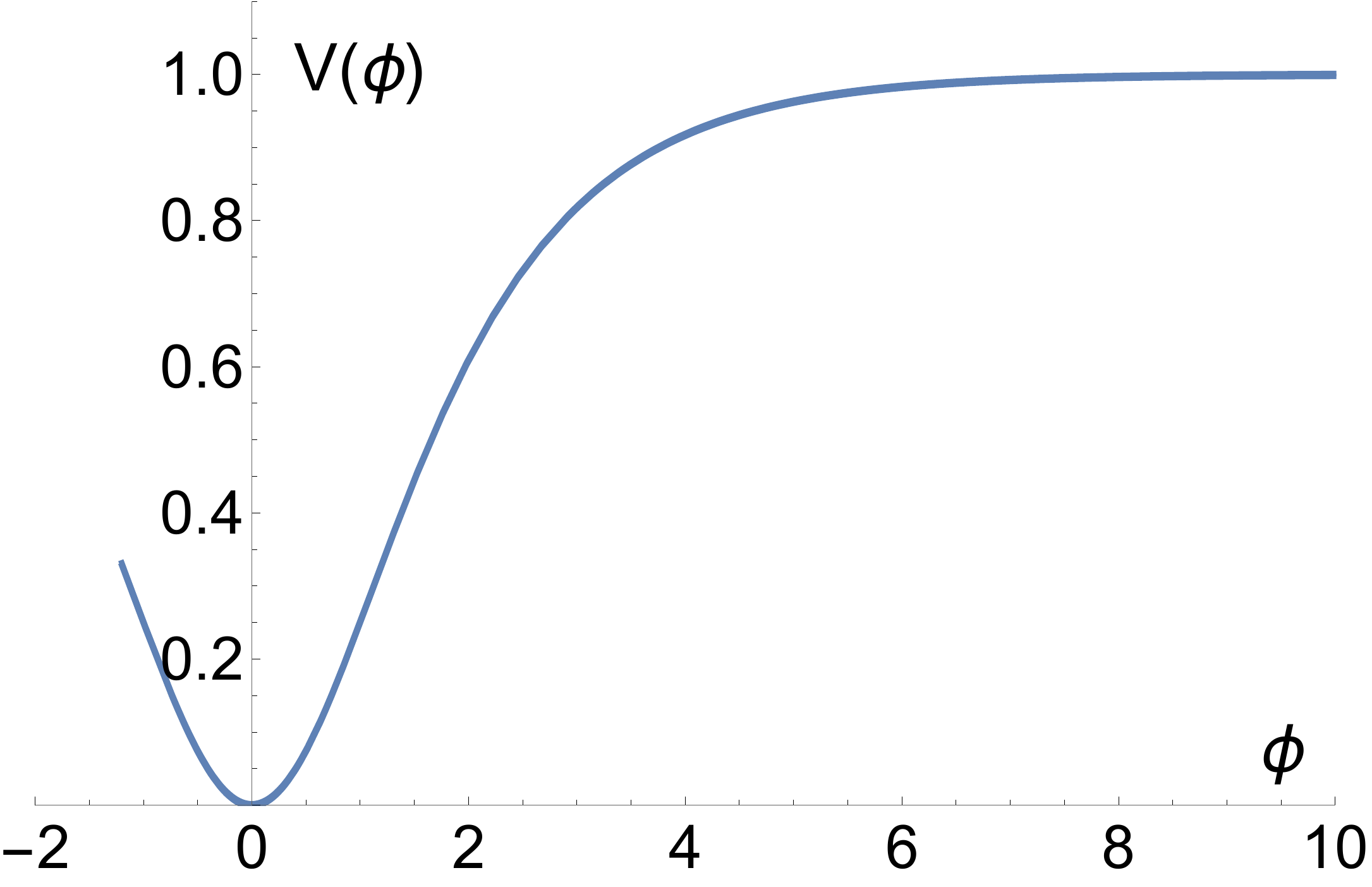}
\caption{\small Schematic plot of the mutated hilltop potential given by Eq.~(\ref{MHI}) as a function of $\phi$ for an inflaton field rolling from the right.}
\label{potmu}
\end{figure}

The mutated hilltop inflation (MUT) model of Pal, Pal and Basu  is given by the potential \cite{Pal:2009sd}, \cite{Pal:2010eb}
\begin{equation}
\label{MHI}
V= V_0 \left(1- \sech(\frac{\phi}{\mu})\right),
\end{equation}
and shown in Fig.~\ref{potmu}.
For this model we can proceed as before however the equations obtained are very complicated and no analytical solution can be found. Instead we start by writing Eq.~(\ref{IA}) as 
\beq
\label{rmu}
r - \frac{2}{3\pi^2 A_s}V= 0\,,
\eeq
from where it follows
\beq
\label{potmusol}
\sech\left(\frac{\phi_k}{\mu}\right) = 1 - \frac{3\pi^2 A_s}{2V_0}r \,.
\eeq
Apparently using Eq.~(\ref{IA}) complicates matters because it introduces the overall constant $V_0$ into the game, which is not the case when using Eqs.~\eqref{Int} and \eqref{Ins}. However, as we will see, following this path it is possible to find an analytical solution. From Eq.~\eqref{Int} 
\beq
\label{reqmu}
r - 16\epsilon = r - \frac{8\sech^2(\frac{\phi_k}{\mu})\tanh^2(\frac{\phi_k}{\mu})}{\mu^2\left(1-\sech(\frac{\phi_k}{\mu})\right)^2}= 0\,,
\eeq
we now solve for $\mu^2$ 
\beq
\label{rmusol}
\mu^2 =  \frac{2(4V_0 - 3\pi^2 A_s r)(3\pi^2 A_s r -2V_0)^2}{3\pi^2 A_s V_0^2 r^2}\,.
\eeq
From Eq.~(\ref{Ins})
\beq
\label{nseqmu}
\delta_{n_s}  + \frac{2\left(2-\cosh^2(\frac{\phi_k}{\mu})\right)\sech^3(\frac{\phi_k}{\mu})}{\mu^2 \left(1-\sech(\frac{\phi_k}{\mu})\right)}-\frac{3}{8}r = 0\,,
\eeq
and substituting Eqs.~(\ref{potmusol}) and (\ref{rmusol}) into Eq.~(\ref{nseqmu}) we can now solve for $V_0$
\beq
\label{v0}
V_0 = \frac{3\pi^2 A_s r\left( r - 24 \delta_{n_s}-\sqrt{17 r^2 +16 r \delta_{n_s} +64 \delta_{n_s}^2}\right)}{16(r-4 \delta_{n_s})}\,,
\eeq
and calculate $V^{\prime \prime \prime }/V^{\prime}$ with the result
\begin{equation}
\frac{V^{\prime \prime \prime }}{V^{\prime}}=  -\frac{(27 r -8 \delta_{n_s})(r + 8 \delta_{n_s})-(13 r+ 8 \delta_{n_s})\sqrt{17 r^2 +16 r \delta_{n_s} +64 \delta_{n_s}^2} }{64 r}\;.
\label{v3vpmu}
\end{equation}
Finally the running $n_{sk}$ can be written as follows
\begin{equation}
n_{sk} = \frac{1}{256}\left(51 r^2 + 80 r  \delta_{n_s} - 64 \delta_{n_s}^2-(13 r + 8  \delta_{n_s})\sqrt{17 r^2 +16 r \delta_{n_s} +64 \delta_{n_s}^2}\right) \;.
\label{nsksolmu}
\end{equation}
\begin{figure}[tb]
\includegraphics[width=10cm]{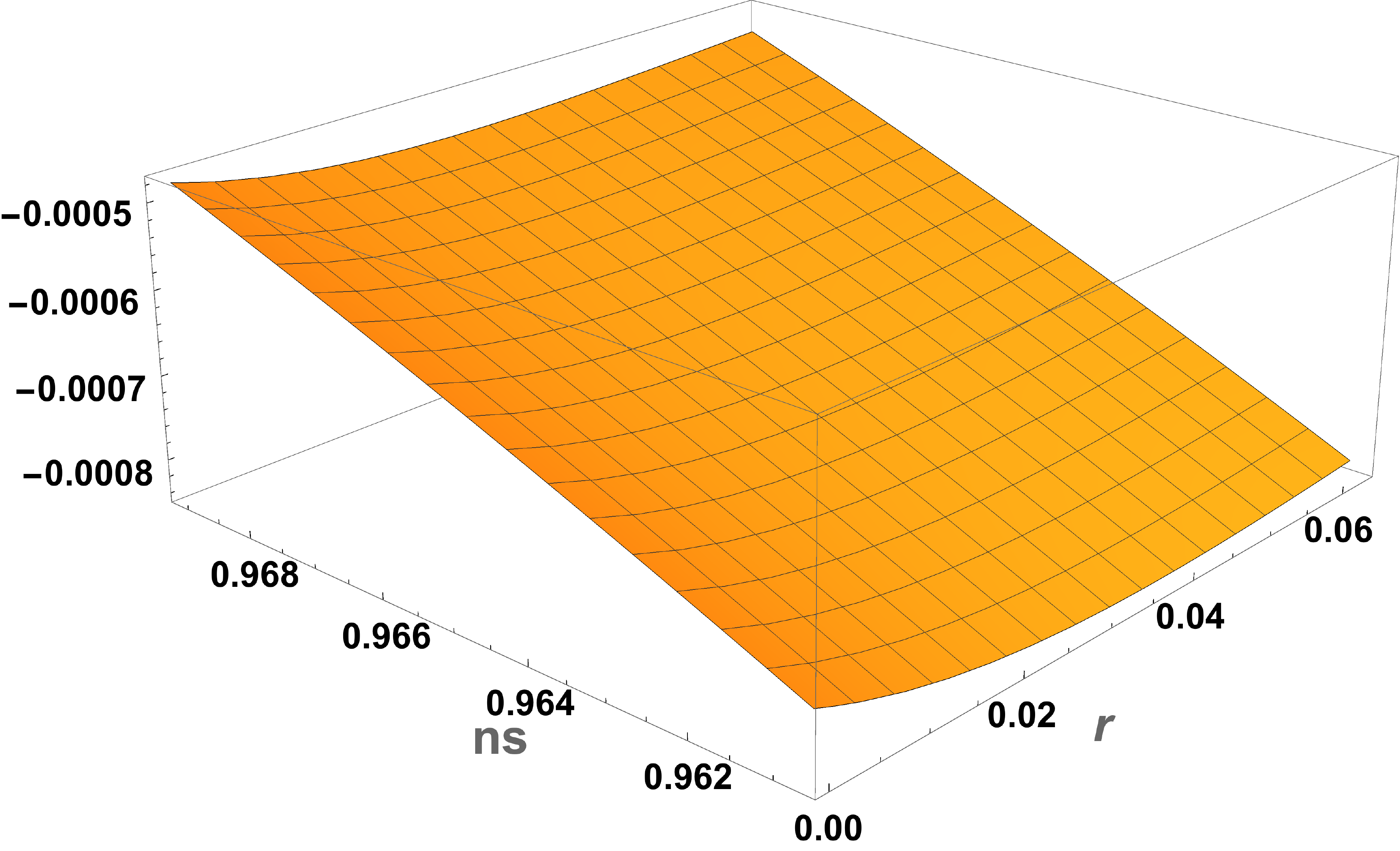}
\caption{\small The running index given by Eq.~(\ref{nsksolmu}) for the mutated hilltop potential of Eq.~(\ref{MHI}). The running is given purely in terms of the observables $n_s$ and $r$ and is plotted for the intervals  $0.9607<n_s<0.9691$ and  $0 > r > 0.063$ given by Planck $2018$, with the resulting bound for $n_{sk}$ as follows: $-8.4\times 10^{-4} < n_{sk} < -5.0\times 10^{-4}$.
}
\label{nskmut}
\end{figure}
This expression is plotted in Fig.~\ref{nskmut} and the bounds are given by
\begin{equation}
0.9607<n_s<0.9691, \quad  0.063 > r > 0, \quad \Rightarrow  \quad -8.4\times 10^{-4} < n_{sk} < -5.0\times 10^{-4}\;.
\label{mutbounds}
\end{equation}
\section {\bf The AFMT model}\label{AFMT}
Here, we apply the results discussed in the previous sections to the AFMT set of models given by the potential \cite{Antusch:2020iyq}
\begin{equation}
\label{potaf}
V(\phi,X)= \frac{1}{p}\Lambda^4 \tanh^p\left(\frac{| \phi|}{M}\right)+\frac{1}{2}g^2\phi^2X^2,
\eeq
where the first term is the inflationary potential (see Fig.~(\ref{potafmt})) and the second gives the interaction of the inflaton with a light field $X$ to which energy is transferred. The parameters $M$ and $\Lambda$ are mass scales, $g$ is a dimensionless coupling and $p$ labels the models.
\begin{figure}[tb]
\includegraphics[width=8cm]{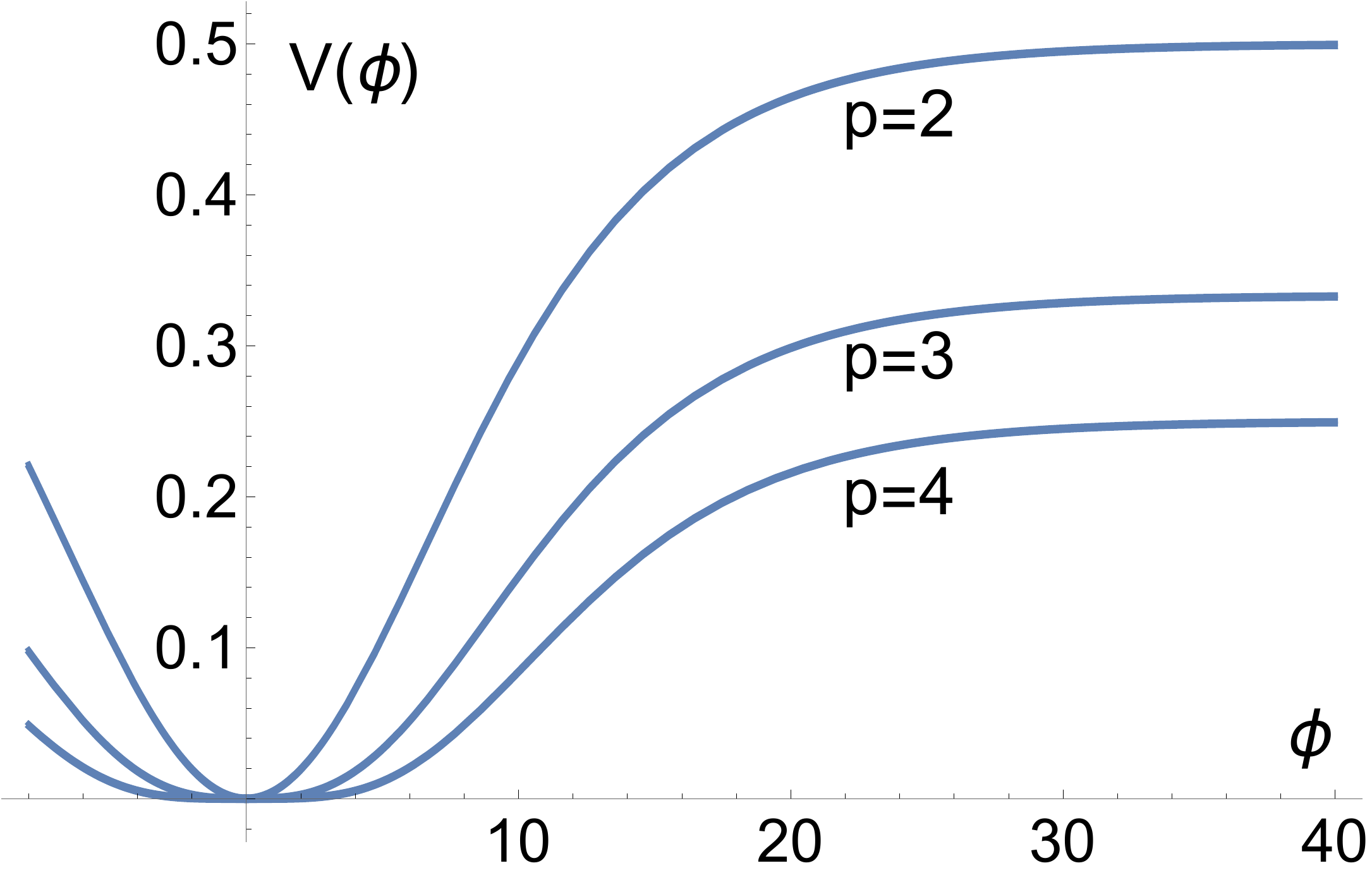}
\caption{\small Schematic plot of the AFMT potential given by Eq.~(\ref{potaf})  as a function of $\phi$ for an inflaton field rolling from the right. The curves correspond to three possible models depending on the parameter $p$ as shown. }
\label{potafmt}
\end{figure}
The first term is what concern us here, the expression $r=16\epsilon$ can be written as
\begin{equation}
r - \frac{8\,p^2\sech^4(\frac{|{\phi_k}}{M})}{M^2\left(1 - \sech^2(\frac{|{\phi_k}|}{M})\right)} = 0\;,
\label{reqafmt}
\end{equation}
from where we get the solution
\begin{equation}
\sech\left(\frac{|{\phi_k}|}{M}\right) = \frac{1}{4\,p}\sqrt{-M^2r+M\sqrt{r}\sqrt{32\,p^2+M^2r}}\;.
\label{reqsolafmt}
\end{equation}
Evaluating  $\eta$ with this solution we find that Eq.~\eqref{Ins} 
\begin{equation}
\delta_{n_s} + \frac{2p\left((1+p)\sech^2\left(\frac{|{\phi_k}|}{M}\right)-2\right)\sech^2\left(\frac{|{\phi_k}|}{M}\right)}{M^2\left(1-\sech^2\left(\frac{|{\phi_k}|}{M}\right)\right)} - \frac{3}{8}r= 0\;,
\label{nseqafmt}
\end{equation}
becomes an equation for $M=M(n_s,r,p)$  with the solution
\begin{equation}
M=  \frac{8\,p\sqrt{2r}}{\sqrt{\left((p-2)r-8\,p\,  \delta_{n_s}\right)\left((p+2)r-8\,p\,  \delta_{n_s}\right)}}\;.
\label{Msolafmt}
\end{equation}
\begin{figure}[tb]
\includegraphics[width=10cm]{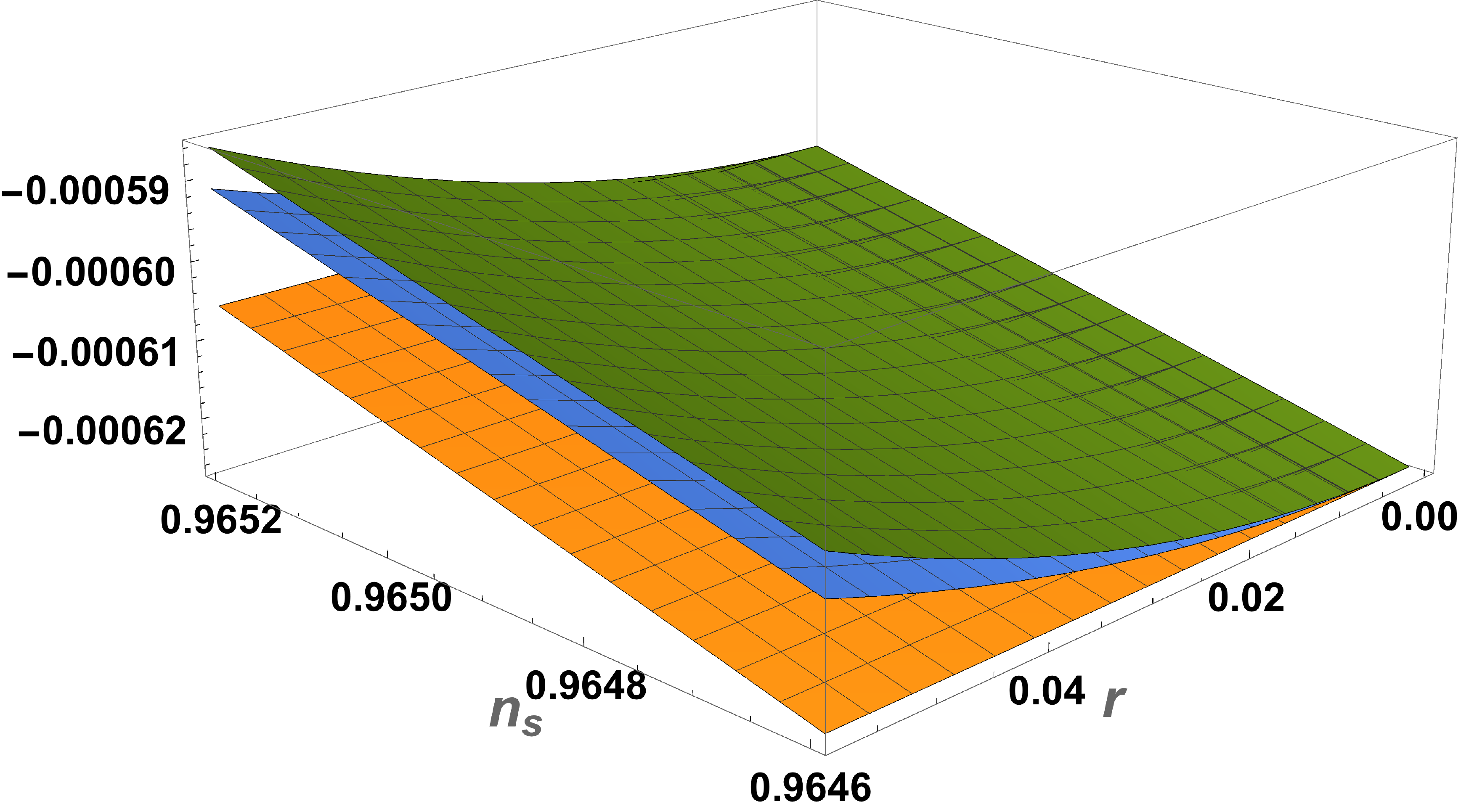}
\caption{\small Plot of the running of the scalar index given by Eq.~\eqref{nskafmt} for the AFMT set of models defined by the potential Eq.~(\ref{potaf}) for three values of $p$: $p=2$ blue (central) sheet, $p=3$ orange (bottom) sheet and $p=4$. Only a section around the central $n_s$ value, from $n_s=0.9646$ to $n_s=0.9652$, is plotted to show the separation of the sheets, eventually converging to $ -7.7\times 10^{-4}$ for $r=0$.
The plotted running is certainly model dependent (labeled by $p$) but does not depends on the parameters $\Lambda$, $M$ of the model itself. The bounds on $n_{sk}$ are given in the Table \ref{tablecomp} and for each $p$ depend exclusively on the phenomenological bounds for the spectral index $n_s$ and tensor-to-scalar ratio $r$ as reported by the Planck collaboration \cite{Aghanim:2018eyx}, \cite{Akrami:2018odb}.
}
\label{nskaf}
\end{figure}
Thus,
\begin{equation}
\frac{V^{\prime \prime \prime }}{V^{\prime}}=  \frac{1}{32} \left(11+\frac{4}{p^2}\right)r-2 \delta_{n_s}+\frac{2\delta_{n_s}^2}{r}\;,
\label{v3vpniafmt}
\end{equation}
this last result together with Eq.~\eqref{nsk}  implies
\begin{equation}
n_{sk}(p) = \frac{1}{128\,p^2}\left(p^2-4\right)r^2- \frac{\delta_{n_s}^2}{2}\;.
\label{nskafmt}
\end{equation}
From the bounds for $n_s$ and $r$, $n_{sk}(p)$ is bounded (for three particular models) as follows 
\begin{equation}
p=2,\quad  -7.7\times 10^{-4} < n_{sk} < -4.8\times 10^{-4}\;,
\label{2bounds}
\end{equation}
\begin{equation}
p=3,\quad  -7.7\times 10^{-4} < n_{sk} < -4.6\times 10^{-4}\;,
\label{3bounds}
\end{equation}
\begin{equation}
p=4,\quad  -7.7\times 10^{-4} < n_{sk} < -4.5\times 10^{-4}\;.
\label{4bounds}
\end{equation}
The lower bound is the same for all cases because, from Eq.~\eqref{nskafmt}, the first term vanishes for $r=0$.
\section {\bf The Starobinsky Model of Inflation}\label{STA}
The Starobinsky model is given by the potential  \cite{Starobinsky:1980te}, \cite{Mukhanov:1981xt} - \cite{Whitt:1984pd}:
\beq
\label{staropot}
V= V_0 \left(1- e^{-\sqrt{\frac{2}{3}}\phi} \right)^2,
\eeq
and is schematically shown in Fig.~\ref{potsta}. 
\begin{figure}[tb]
\includegraphics[width=8cm]{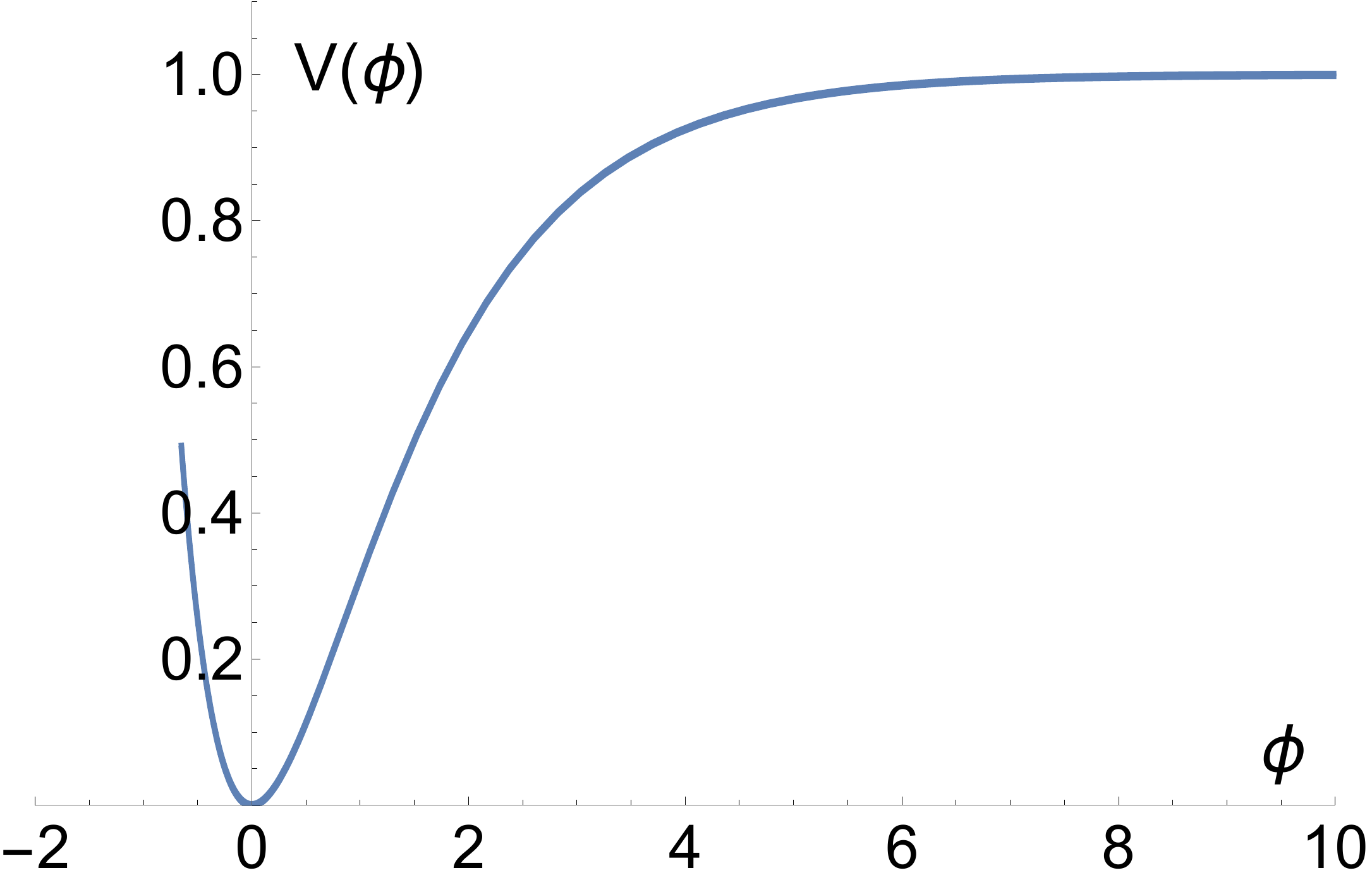}
\caption{\small Schematic plot of the Starobinsky potential given by Eq.~(\ref{staropot}) as a function of $\phi$ for an inflaton field rolling from the right. }
\label{potsta}
\end{figure}
This is a one-parameter model, being $V_0$ an overall constant it does not appear in Eqs.~(\ref{Int}) and (\ref{Ins}). Thus, we can obtain the solution for $\phi_k$ directly in terms of $n_s$ by solving Eq.~(\ref{Ins}) or in terms of $r$ by solving Eq.~(\ref{Int}). In the first case we get 
\beq
\label{staphins}
\phi_k= \sqrt{\frac{3}{2}} \ln\left(\frac{4+3\delta_{n_s}+4\sqrt{1+3\delta_{n_s}}}{3\delta_{n_s}} \right),
\eeq
where, as before, $\delta_{n_s}\equiv 1-n_s$. Once we have $\phi_k$ we can calculate any quantity of interest during inflation, in particular
\beq
\label{star}
r=4\delta_{n_s}-\frac{8}{3}\left(\sqrt{1+3\delta_{n_s}}-1\right).
\eeq
From the Planck range for the scalar spectral index $n_s = 0.9649 \pm 0.0042$ we get $r = 0.0036 \pm 0.0008$. 

We can immediately calculate the running index as a function of $n_s$ only
\beq
\label{stanskr}
n_{sk}(n_s)=-\frac{1}{18}\left(2+3\delta_{n_s}-2\sqrt{1+3\delta_{n_s}}\right)\left(6+3\delta_{n_s}-2\sqrt{1+3\delta_{n_s}}+5\sqrt{2+3\delta_{n_s}-2\sqrt{1+3\delta_{n_s} }   }\right),
\eeq
this running index is plotted in Fig.~\ref{nsksta}. From the Planck range for the spectral index given above we get $n_{sk} = -(6.3 \pm 1.5)\times 10^{-4}$, also $n_{tk} = -(1.6 \pm 0.5)\times 10^{-5}$.
\begin{figure}[tb]
\includegraphics[width=10cm]{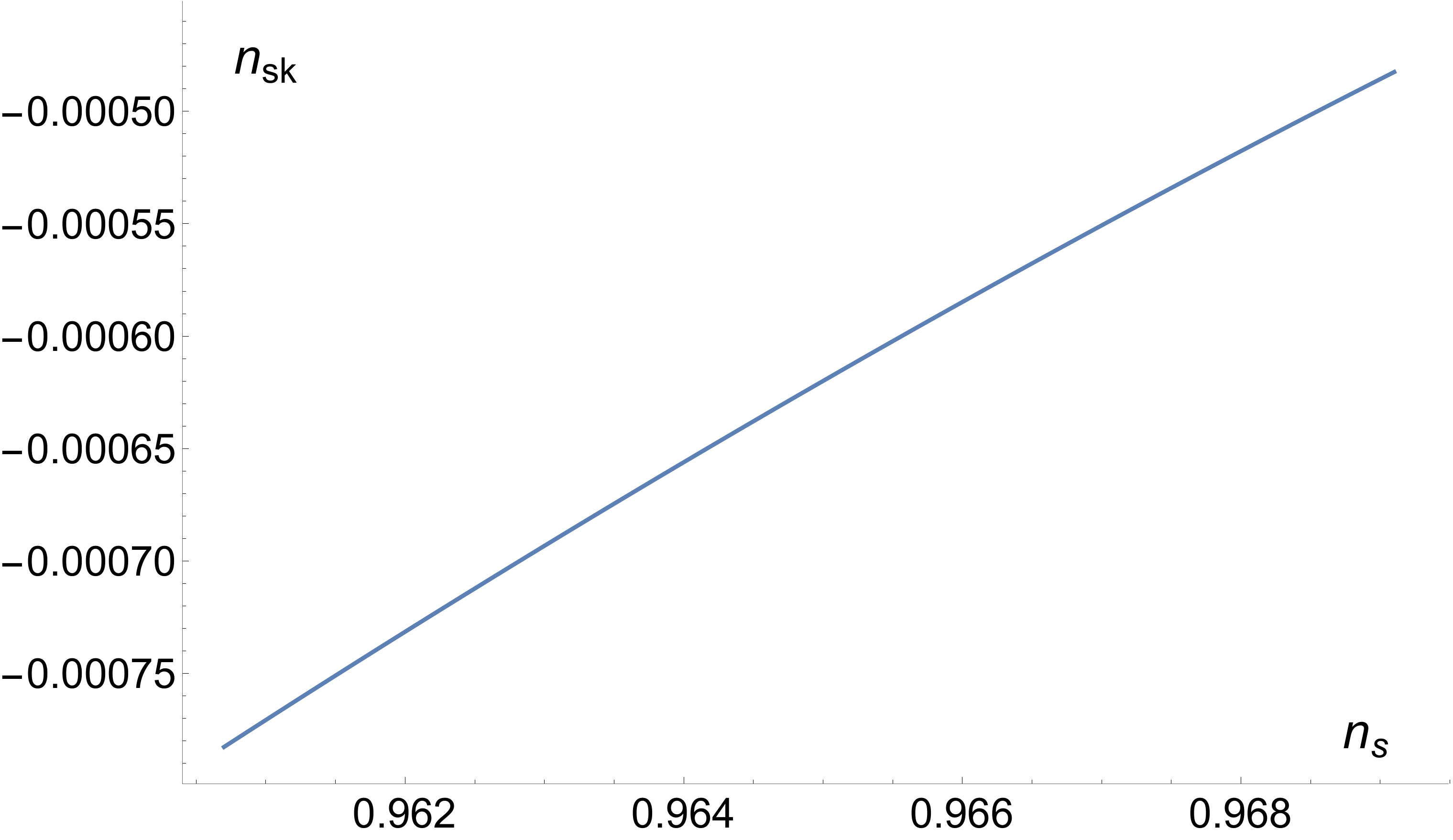}
\caption{\small In the Starobinsky model we can solve Eq.~(\ref{Ins}) for $\phi$ as function of $n_s$ as given by Eq.~(\ref{staphins}) or Eq.~(\ref{Int}) for $\phi$ as a function of $r$ as given by Eq.~(\ref{fista}). In the first case the running is plotted above and given by Eq.~(\ref{stanskr}) with the bounds $-7.8\times 10^{-4} < n_{sk} <-4.8\times 10^{-4}.$ The bounds coming from $n_s$ also impose the following bounds on $r:$\, $0.0028 < r < 0.0044$. 
}
\label{nsksta}
\end{figure}
 \begin{center}
\addtolength{\tabcolsep}{-7pt}
\begin{table*}[htbp!]
\caption{\label{tablecomp} In the table below we collect the bounds for the various running indices  obtained in the article. The range of values for the scalar spectral index $n_s$ and the tensor-to-scalar ratio $r$ is the one given by the Planck Collaboration: $0.9607 < n_s < 0.9691$ and $0 < r < 0.063$, respectively. The only exception occurs for the Starobinsky model where $r$ can be expressed in terms of $n_s$ as shown in Eq.~(\ref{star}) thus, $r$ is constrained by $n_s$ to the range $0.0028 < r < 0.0044$. The resulting bound for the running of the tensor index $n_{tk}$ is model independent: $-2.47\times 10^{-4} < n_{tk} < 0$ (see Eq.~(\ref{ntkcons})) and it should be satisfied by any single field model of inflation. Planck $2018$ sets $n_{sk}=-0.005\pm 0.013$ at $95\%$ CL with no running of the running being considered.
}
\begin{ruledtabular}
\begin{tabular}{ccc}
$Model$ & $V(\phi)$ & $\Delta n_{sk}$\\ \hline\\[0.1mm]
NI   &  $V(\phi) =V_0\left(1-\cos(\frac{\phi}{f})\right)$,\,\,\,\,\,\,\,Eq.~(\ref{nipot})  & $-4.9\times 10^{-4} < n_{sk} < 0$\\[2mm] 
MHI   &  $V(\phi)= V_0 \left(1- \sech(\frac{\phi}{\mu})\right)$,\,\,\,\,Eq.~(\ref{MHI}) & $-8.4\times 10^{-4} < n_{sk} < -5.0\times 10^{-4}$\\[2mm]
AFMT(p=2)  &   $V(\phi)= \frac{1}{p}\Lambda^4 \tanh^p\left(\frac{|\phi|}{M}\right)$,\,\,\,\,\,\,\,Eq.~(\ref{potaf}) & $-7.7\times 10^{-4} < n_{sk} < -4.8\times 10^{-4}$\\[2mm]
AFMT(p=3)   &  $--$ & $-7.7\times 10^{-4} < n_{sk} < -4.6\times 10^{-4}$\\[2mm]
AFMT(p=4)   &  $--$ & $-7.7\times 10^{-4} < n_{sk} < -4.5\times 10^{-4}$\\[2mm]
Starobinsky  & $V(\phi)= V_0 \left(1- e^{-\sqrt{\frac{2}{3}}\phi} \right)^2$,\,\,\,Eq.~(\ref{staropot}) & $-7.8\times 10^{-4} < n_{sk} < -4.8\times 10^{-4}$\\[2mm] 
\end{tabular}
\end{ruledtabular}
\end{table*}
\end{center}
We can also solve Eq.~(\ref{Int}) 
\beq
\label{rsta}
r - \frac{64}{3\left(e^{\sqrt{\frac{2}{3}}\phi_k}-1 \right)^2} = 0,
\eeq
for $\phi_k$ in terms of $r$ with the result
\beq
\label{fista}
\phi_k = \sqrt{\frac{3}{2}}\ln\left(1+\frac{8}{\sqrt{3r}} \right).
\eeq
The running can now be written more economically in terms of $r$ 
\beq
\label{nskrsta}
n_{sk}(r)=-\frac{1}{96}r\left(16+10\sqrt{3r} +3r\right).
\eeq
To first order in $r$ the consistency relations given by Eqs.~(\ref{star}) and (\ref{nskrsta}) above reduce to Eq.~(32) of Ref. \cite{Motohashi:2014tra} where $R^2$ and its generalization is studied in detail. 

Finally we would like to remark that although all the expressions for $n_{sk}$ of the  models studied are  different this is not enough to believe that, in general, the running could break the degeneracy of models of inflation. As a counterexample  let us consider the central value of $n_s$ as reported by Planck i.e., $n_s=0.9649$, it is not difficult to show the the MHI model for $\mu \approx 1.24$ gives $r=0.0035$ and $n_{sk}=-0.0006$ exactly the same values (at this level of approximation) to the ones obtained from the Starobinsky model. To break the degeneracy it could be necessary to go to the reheating epoch where important differences between models can arise  \cite{German:2020iwg}.
In Table \ref{tablecomp} we compare results for the running index for the NI, MHI, ATMF and Starobinsky models of inflation.

We close our article with a final consideration. As we can see from the Table \ref{tablecomp}, $n_{sk}$ is of $\cal O$$(10^{-3}-10^{-4})$ and if we calculate the running of the running $n_{skk}\equiv d^2 n_s/d\ln k^2$ along the same line of arguments as before we will find that it is of $\cal O$$(10^{-5})$. Thus, this quantities, compared to $n_s$, are small indeed and in certain circumstances they could be neglected however, one should be careful not to make them zero. From the general expression for $n_{sk}$ given by Eq.~\eqref{nsk} we see that making $n_{sk}=0$ implies
\begin{equation}
\frac{V^{\prime \prime \prime }}{V^{\prime}} = \frac{1}{8}(3r-16\delta_{n_s})\;,
\label{nsk2}
\end{equation}
which cannot possibly be true in general (none of the models studied before have this expression for $V^{\prime \prime \prime }/V^{\prime}$).
An equivalent way of seen this is through the explicit solutions for  $n_{sk} = 0$, taking NI as an example $n_{sk} = \frac{1}{32}r\left(r-8 \delta_{n_s}\right)=0$ would imply the solutions $r=0$ and/or $r=8 \delta_{n_s}$ none of them consistent with the bounds for $r$: $0 < r < 0.063$. Something similar occurs for the running of the running of the scalar index which for NI is $n_{skk}=\frac{1}{32}r\left(r-8 \delta_{n_s}\right)\delta_{n_s}$ making $n_{skk}=0$ will give the extra unacceptable solution $\delta_{n_s}=0$ or $n_s=1$. Thus, one must be careful to distinguish between neglecting a term and making it zero.
\section {\bf Conclusions}\label{CON}
Bounds for the running have been studied for both the running of the tensor index $ n_ {tk} $ and the running of the scalar index $ n_ {sk}$ (also denoted $\alpha $) for four inflationary models: Natural Inflation, Mutated Hilltop Inflation, the AFMT model and the Starobinsky model (see Table \ref{tablecomp}). In all cases, the running has been written in terms of the scalar spectral index $ n_s $ and the tensor-to-scalar-ratio $ r $ only without the presence of any parameters of the model in question. This allows to obtain bounds for $ n_ {sk} $ directly from the observables. These bounds will be narrowed by more precise observations and/or by finding other restrictive conditions on $ n_s $ and $r$. 
The problem of the degeneracy of inflationary models has been briefly discussed and the role of the running in breaking such degeneracy could be important however, the running is not in general enough to break the degeneracy of models of inflation pointing towards the study of the reheating epoch to achieve this.

\acknowledgments

We acknowledge financial support from UNAM-PAPIIT,  IN104119, {\it Estudios en gravitaci\'on y cosmolog\'ia}.


\end{document}